\documentclass[conference,10pt]{IEEEtran}
 
\usepackage{datenumber}
\usepackage{amsmath} 
\usepackage{amsfonts} 
\usepackage{verbatim}  
\usepackage{color} 
\usepackage{booktabs} 
\usepackage{rotating}
\def\arraystretch{1.6}
\usepackage{nicefrac} 
\usepackage{bigstrut}     
\usepackage{array}  
\usepackage{arydshln}  
\usepackage{pgfplots} \usetikzlibrary{plotmarks,patterns}
 \usepackage{url}
 \usepackage{enumerate}
\usepackage{mathtools}
\usepackage{balance} 

\usepackage{rotating}

\newcommand{\Span}[1]{\mathsf{span} \!\left(  #1  \right)}
\newcommand{\RSpan}[1]{\mathsf{rspan} \left( #1 \right)}

\newcommand{\Diag}{\mathsf{bdiag}}
\newcommand{\diag}[1]{\Diag \left( #1 \right)}
\newcommand{\Stack}{\mathsf{stack}}
\newcommand{\stack}[1]{\Stack \left( #1 \right)}

\newcommand{\pare}[1] {\left( #1 \right)} 
\newcommand{\parep}[1] {\big( #1 \big)} 
\newcommand{\Min}[1] {\min \pare{#1}}
\newcommand{\Max}[1] {\max \parep{#1}} 
\DeclarePairedDelimiter\ceil{\lceil}{\rceil}

\newcommand{\0}{\mathbf{0}}
\newcommand{\Ac}{\mathcal{A}}
\newcommand{\An}{\mathbf{A}}

\newcommand{\BS}[1]{\text{BS}_{#1}}

\newcommand{\Cmat}[2]{\!\in\mathbb{C}^{#1 \times #2}}

\newcommand{\djin}{d_j^{(\text{in})}}

\newcommand{\enters}{\mathbb{Z}^+}

\newcommand{\Gn}{\mathbf{G}}
 
\newcommand{\Hn}{\mathbf{H}}
\newcommand{\hn}{\mathbf{h}}
\newcommand{\hns}[2]{\underline{\hn}_{#1}^{\left(#2\right)}} 
\newcommand{\Hns}[2]{\Hn_{#1}^{\left(#2\right)}}
\newcommand{\Hnsg}[2]{\bar{\Hn}_{#1}^{\left(#2\right)}}

\newcommand{\inn}{\mathbf{i}}

\newcommand{\Ic}{\mathcal{I}}

\newcommand{\roA}{\rho_{\text{A}}}
\newcommand{\roB}{\rho_{\text{B}}}
\newcommand{\roC}{\rho_{\text{C}}}
\newcommand{\sigmas}[2]{\boldsymbol{\theta}_{#1}^{(#2)}}
\newcommand{\Sigmas}[2]{\boldsymbol{\Theta}_{#1}^{(#2)}}

\newcommand{\sn}{\mathbf{s}}
\newcommand{\tng}[2]{\underline{\un}_{#1}^{(#2)}}
\newcommand{\Tc}{\mathcal{T}}
\newcommand{\Tn}[1]{\mathbf{T}_{#1}}

\newcommand{\un}{\mathbf{u}}

\newcommand{\UE}[1]{\text{UE}_{#1}}

\newcommand{\Vn}{\mathbf{V}}

\newcommand{\Vns}[2]{\Vn_{#1}^{\left(#2\right)}}

\newcommand{\wn}{\mathbf{w}}
\newcommand{\Wn}[1]{\underline{\wn}_{#1}}

\newcommand{\xn}{\mathbf{x}}
\newcommand{\Xn}{\mathbf{X}}
\newcommand{\Xc}{\mathcal{X}}
\newcommand{\yn}{\mathbf{y}}
\newcommand{\Yn}{\mathbf{Y}}
\newcommand{\Yc}{\mathcal{Y}}

\newcommand{\sigmaU}[8]{\! \Big[ \sigmas{#1}{#2}\!,\sigmas{#3}{#4} \Big] \!\!\! \begin{bmatrix}\tng{#5}{#6} \\ \tng{#7}{#8} \end{bmatrix}}

\newtheorem{theorem}{Theorem}

\IEEEoverridecommandlockouts

\newcolumntype{C}{>{$}c<{$}} 
\newcolumntype{L}{>{$}l<{$}}
\newcolumntype{R}{>{$}r<{$}}

\definecolor{creme}{RGB}{255, 253, 208}

\newcommand{\poligon}[1]{\fill [color=creme!90, postaction=
    {pattern=north east lines, pattern color = black!15}] #1;}

\author{ Marc Torrellas, Adrian Agustin, and Josep Vidal \\[3mm]
{Signal Theory and Communications Department} \\ Universitat Polit\`ecnica de Catalunya (UPC), Barcelona \\ \{marc.torrellas.socastro, adrian.agustin, josep.vidal\}@upc.edu
\thanks{{This work has been supported by projects TROPIC FP7 ICT-2011-8-318784 (European Commission), MOSAIC TEC2010-19171/TCM, DISNET TEC2013-41315-R (Ministerio de \mbox{Economia} y Competitividad, Spanish Government and ERDFs), and \mbox{2014SGR-60} (Catalan Administration).}}
}

\title{Retrospective Interference Alignment \\ for the MIMO Interference Broadcast Channel \\[-2.2mm]}

\begin{document}

\begin{figure*}
Paper submitted to IEEE International Symposium on Information Theory (ISIT) 2015.
\\[2mm]
\copyright 2015 IEEE. Personal use of this material is permitted. Permission from IEEE must be 
obtained for all other uses, in any current or future media, including 
reprinting/republishing this material for advertising or promotional purposes, creating new 
collective works, for resale or redistribution to servers or lists, or reuse of any copyrighted 
component of this work in other works. 
 \end{figure*}

\newpage

\maketitle

%
%

\begin{abstract}
The degrees of freedom (DoF) of the multiple-input multiple-output (MIMO) Interference Broadcast Channel (IBC) with 2 cells and 2 users per cell are investigated when only delayed channel state information is available at the transmitter side (delayed CSIT). Retrospective Interference Alignment has shown the benefits in terms of DoF of exploiting delayed CSIT for interference, broadcast and also for the IBC. However, previous works studying the IBC with delayed CSIT do not exploit the fact that the users of each cell are served by a common transmitter. This work presents a four-phase precoding strategy taking this into consideration. Assuming that transmitters and receivers are equipped with $M,N$ antennas, respectively, new DoF inner bounds are proposed, outperforming the existing ones for $\rho = \frac MN> 2.6413$.
\end{abstract}

\begin{IEEEkeywords}
MIMO, Interference Alignment, Degrees of freedom, Delayed CSIT, Interference Broadcast Channel.
\end{IEEEkeywords}

\section{Introduction}

%
%

Channel capacity characterization remains still unknown for many wireless interference network settings. In this regard, the study of channel degrees of freedom (DoF) has become a crowded research avenue during the last decade. The DoF represents the scaling of capacity with respect to the signal-to-noise ratio (SNR) at the high SNR regime, and allows to elucidate the impact of different antenna settings or feedback assumptions on channel capacity. One of the key elements in this context is the emergence of interference alignment (IA), a new strategy for managing the signal dimensions (time, frequency, space) in pursuit of DoF maximization \cite{CJ}. The concept consists in designing the transmitted signals in such a way that they are overlapped (or \textit{aligned}) at the non-intended receivers. Therefore, the interference lies on a reduced dimensional subspace, thus releasing some dimensions to allocate desired signals. 
A very extensive survey {of IA applications} can be found in \cite{IAtutorial}.

\begin{figure}[] 
  \centering 
  \centerline{\includegraphics[width=0.9\linewidth]{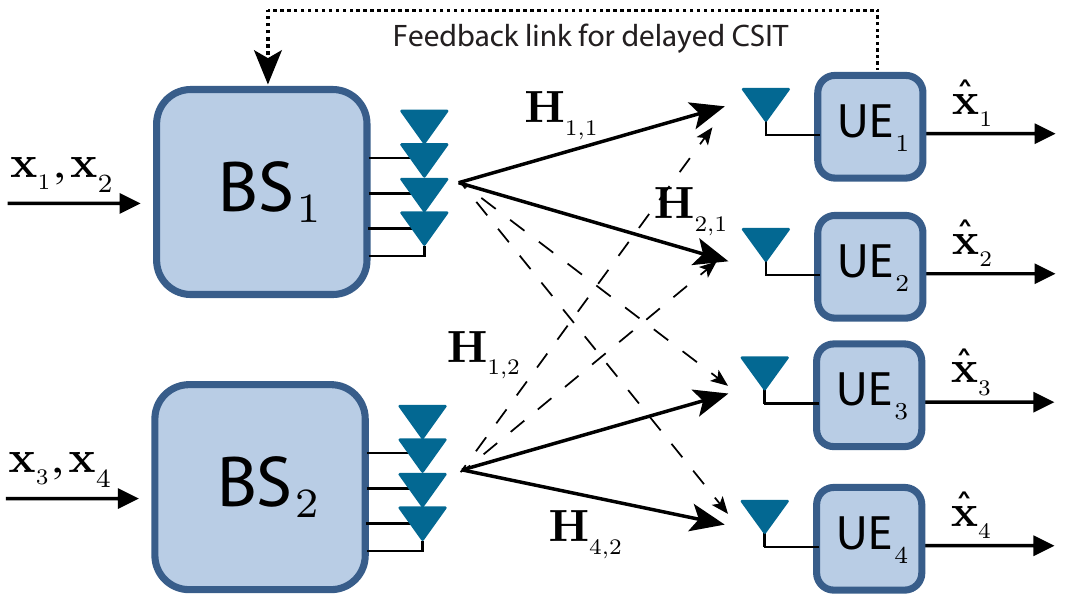}} 
\caption{The IBC with delayed CSIT, 2 cells, 2 users per cell, $M=4$ antennas at the BSs and single-antenna ($N=1$) UEs. Solid lines denote the direct links, providing intended signals, while dashed lines denote the cross links, providing interfering or non-desired signals.}
\vspace{-5mm}
\label{fig:scenario}
\end{figure} 

Usually, IA schemes require current channel state information at the transmitter side (CSIT), an assumption not always valid in wireless networks, especially when channel coherence time is low. Recently, Maddah-Ali and Tse (MAT) introduced in \cite{MAT} for the broadcast channel (BC) a new framework where IA concepts can be exploited even when the CSIT is perfect but completely outdated, referred to as delayed CSIT. The publication of \cite{MAT} inspired many works that extended these ideas to Interference channel (IC) and X channel (IC) \cite{Vaze2IC}\cite{rankRatioMIMOXC}\cite{Abdoli_IC}. 

Initially proposed for the MISO BC, the signals are aligned over the space-time domain by developing a transmission along multiple phases,
and exploiting linear beamforming and user scheduling. 
During the first phase, the base-station (BS) sends linear combinations (LCs) of all symbols of each user in a TDMA fashion, i.e. first the symbols of user equipment (UE) one ($\UE{1}$), then symbols of $\UE{2}$, and so on. In this case, each UE observes less LCs than the number of symbols, thus the symbols cannot be linearly decoded yet. However, due to the randomness of wireless channels, different and independent LCs of each set of symbols are observed at each UE. Those LCs are known at one user UE, and desired at another. Therefore, they may be retransmitted during next phases in such a way that they can be removed at the non-intended UEs, and provide additional LC of desired symbols to the intended UE. Interestingly, this can be interpreted as an interference alignment strategy acting retrospectively. For this reason, this type of IA is denoted as {\it Retrospective IA}. 

This work investigates the DoF of the 2-user 2-cell $(M,N)$ MIMO IBC with delayed CSIT, see Fig. \ref{fig:scenario}, where BSs and UEs are equipped with $(M,N)$ antennas, respectively. This scenario can be modeled as a 4-user $(M/2,N)$ IC where two pairs of BSs cooperate. The first results assuming delayed CSIT were provided by \cite{IBC_dCSIT_ISIT14} for $N=1$ and $M=2,3,4$, but their scheme does not exploit the benefits of having two cooperating transmitters per cell. Consequently, their results basically consist in applying the techniques for the IC, reported in \cite{TAV}.
This work proposes a four-phase transmission for the specific topology of the IBC, taking into account that the users of each cell are served by the same BS. The achieved DoF outperform any previous work for $\rho = \frac{M}{N}>\roA\approx 2.6413$.

\subsection{Notation} 

Boldface and lowercase types denote column vectors ($\mathbf{x}$), while row vectors are underlined ($\underline{\mathbf{x}}$). Boldface and uppercase types are used for matrices ($\mathbf{X}$). Sets and subspaces are denoted by uppercase calligraphic types ($\mathcal{X}$). 

We define $\0$ and $\mathbf{I}$ as the all-zero and identity matrices, respectively, with suitable dimensions according to the context. For vectors and matrices, $(\cdot)^T$ and $(\cdot)^H$ are the transpose and transpose and conjugate operators, respectively. Moreover, the following two predefined operations are defined:
\begin{IEEEeqnarray*}{c c c}
\def\arraystretch{1.2}
\stack{\Xn,\Yn} =
\begin{bmatrix}
\Xn \\
\Yn \\
\end{bmatrix},
& \quad &
\diag{\Xn,\Yn} =
\begin{bmatrix}
	\Xn & \0 \\
	\0 & \Yn \\
\end{bmatrix}.
\end{IEEEeqnarray*}
$\Span{\Xn}$ is usually used to define the \textit{column subspace}, containing all possible linear combinations (LCs) of the columns of $\xn$. However, in this work we use the \textit{row subspace}, defined as ${\Xc = \RSpan{\Xn}=
\mathsf{span} \big( \Xn^T \big) }$. 
Finally, the set $\Xc \backslash \Yc$ contains the elements that belong to $\Xc$ but not to $\Yc$.

\section{System model}
\label{sec:systModel}

\begin{figure}[] 
  \centering 
  {\includegraphics[width=0.95\linewidth]{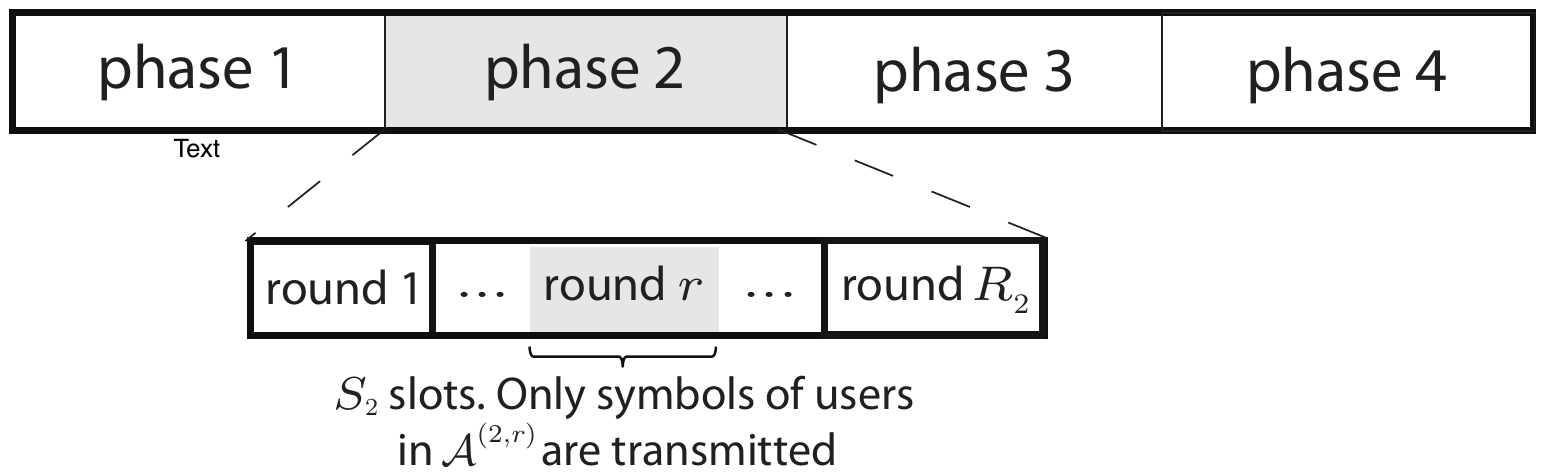}} 
\caption{{General structure of the transmission frame. There are four phases. Phase $p$ is divided in $R_p$ rounds, in turn divided in $S_p$ time slots.} }\vspace{-4mm}
\label{fig:frame} 
\end{figure}

We consider the 2-cell 2-user IBC, see Fig. \ref{fig:scenario}. In this scenario, two BSs equipped with $M$ antennas transmit independent messages to each of its two associated UEs, equipped with $N$ antennas each. UEs are labeled from 1 to 4, thus $\BS{c(i)}$ serves $\UE{i}$, with $c(i)=\ceil{\nicefrac{i}{2}}$.
Communication is carried out in a frame consisting of $P=4$ phases of duration $R_p$ rounds, in turn divided in $S_p$ time slots each, see Fig. \ref{fig:frame}. The total number of slots used for data transmission is denoted by
\begin{IEEEeqnarray}{c}
 \tau=\sum_{p=1}^{P} \tau_p ,  \quad \tau_p = R_p S_p, \quad R_p = \binom{4}{p}.
\end{IEEEeqnarray}
The $(p,r)$th round, i.e. round $r$ of phase $p$, is dedicated to a specific group of $p$ users (served UEs), denoted by $\Ac^{(p,r)}$. The users not in $\Ac^{(p,r)}$ are denoted as the {\it listening users}. The particular users in each $\Ac^{(p,r)}$ may be found in the second row of Table \ref{tab:OHI}. Note that $R_p$ corresponds to the number of possible $p$-tuples that can be formed, {and for simplicity, they are ordered by selecting first the users with the lowest index.} 

The output at the $j$th receiver during the slots of the $(p,\!r)$th round is described by:
\\[-4mm]
\begin{IEEEeqnarray}{c} 
\yn_j^{\left(p,r\right)} =\! \sn_j^{\left(p,r\right)}\! + \inn_j^{\left(p,r\right)} \!=\!
\Hns{j,c(j)}{p,r} \Vns{j}{p,r} \mathbf{x}_j + \sum_{\substack{i=1 \\ i\neq j}}^4 \Hns{j,c(i)}{p,r} \Vns{i}{p,r} \mathbf{x}_i \nonumber \\[-5mm]
\label{eq:SystemModel}
\end{IEEEeqnarray}
\\[-3mm]
where $\yn_j^{\left(p,r\right)} \Cmat{NS_p}{1}$ is the received signal at $\UE{j}$ during the $(p,r)$th round, $\mathbf{x}_i \Cmat{b}{1}$ contains the $b$ uncorrelated unit-powered complex-valued data symbols intended to the $i$th receiver, $\Vn_i^{\left(p,r\right)}\Cmat{MS_p}{b}$ is the precoding matrix carrying the signals intended to $\UE{i}$, designed subject to a maximum transmission power constraint, and with ${\Vn_i^{\left(p,r\right)}=\0,\forall i \notin \Ac^{(p,r)}}$. Note that  
since the focus of this paper is on DoF analysis, all noise terms are omitted.

The channel gains for each slot and each link between transmitter and receiver are described by an $N\times M$ matrix. Then, the channel matrix $\Hn_{j,i}^{\left(p,r\right)} \Cmat{NS_p}{MS_p}$ in (\ref{eq:SystemModel}) is formed as the block diagonal composition of $S_p$ of such matrices, thus contains the channel gains from antennas of $\BS{i}$ to $\UE{j}$ during all time slots of the $(p,r)$th round. A flat block fading channel model is assumed, i.e. channels are i.i.d. as $\mathcal{CN}\!\pare{0,1}$, and completely uncorrelated in time and space.

\begin{table*}
\caption{Served UEs ($\Ac^{(p,r)}$) and available overheard interference (OHI) at $\UE{1}$ for each round.}
\hspace{-2mm}
\scalebox{0.87}{
\def\arraystretch{2}
\begin{tabular}{|C | C | C | C | C | C | C | C | C | C | C | C | C | C | C |}
\hline
\multicolumn{4}{|c|}{Phase 1} & \multicolumn{6}{c|}{Phase 2} &
\multicolumn{4}{c|}{Phase 3} & \multicolumn{1}{c|}{Phase 4} \\
\hline 
 \rotatebox{90}{ \hspace{2.5mm}\rotatebox{-90}{  \{1\} }} & 
 \rotatebox{90}{ \hspace{2.5mm}\rotatebox{-90}{  \{2\} }} & 
 \rotatebox{90}{ \hspace{2.5mm}\rotatebox{-90}{  \{3\} }} & 
 \rotatebox{90}{ \hspace{2.5mm}\rotatebox{-90}{  \{4\} }} & 
\rotatebox{90}{ \,\{1,2\} }  & 
\rotatebox{90}{ \,\{1,3\} } &
\rotatebox{90}{\,\,\,\{1,4\} } &
\rotatebox{90}{\,\,\,\{2,3\}} &
\rotatebox{90}{\,\,\,\{2,4\}} &
\rotatebox{90}{\,\,\,\{3,4\}} &
\rotatebox{90}{\,\{1,2,3\}} &
\rotatebox{90}{\,\{1,2,4\}} &
\rotatebox{90}{\,\{1,3,4\}} &
\rotatebox{90}{\,\{2,3,4\}} &
\rotatebox{90}{\!\{1,2,3,4\} } \\
\hline
     & \Tn{1,2} & \Tn{1,3} & \Tn{1,4} &
 & & & \Big[ \tng{3,2}{1}, \tng{2,3}{1} \Big] & 
 \Big[ \tng{4,2}{1}, \tng{2,4}{1} \Big] &
 \Big[ \tng{4,3}{1}, \tng{3,4}{1} \Big] & 
 \tng{3,2}{1}, \tng{2,3}{1} &
 \tng{4,2}{1}, \tng{2,4}{1}  & &
 \Big[ \Wn{1,2},\Wn{1,3},\Wn{1,4} \Big] & \\[0.8mm]
\hline
\end{tabular} 
}
\label{tab:OHI}
\end{table*}

After the whole communication, $\UE{j}$ collects all the received signals. Since linear transmitters and receivers are used, the objective is to obtain at least $b$ LCs of desired symbols free of interference at each receiver. In this regard, let write the channel model in (\ref{eq:SystemModel}) in a more compact form by grouping the magnitudes of the different phases and rounds, as follows:  
\begin{IEEEeqnarray*}{c}
\label{eq:SystemModelextended}
\def\arraystretch{1.2}
\yn_j^{(p)} =   \Gn_j^{(p)} \left[\xn_1^T,\dots,\xn_4^T \right]^T = \stack{\yn_j^{(p,1)},\ldots,\yn_j^{(p,R_p)}}, \\
\Gn_j^{(p)} =
    \begin{bmatrix}
		\Hn_{j,1}^{(p)} \Vn_{1}^{(p)}, \, \ldots ,\,
		\Hn_{j,2}^{(p)} \Vn_{4}^{(p)}
	\end{bmatrix}, \\ 
\Hn_{j,i}^{(p)} = \diag{
\begin{matrix}
	\Hn_{j,i}^{\left(p,1\right)}, \,
	\Hn_{j,i}^{\left(p,2\right)}, \,\ldots, \,
	\Hn_{j,i}^{\left(p,R_p\right)}
\end{matrix} }, \\[-0.5mm]
\Vn_{i}^{(p)} = \stack{
\begin{matrix}
	\Vn_{i}^{\left(p,1\right)}, \,
	\Vn_{i}^{\left(p,2\right)}, \,\ldots, \,
	\Vn_{i}^{\left(p,R_p\right)}
\end{matrix} }, \\[-0.5mm]
\Hn_{j,i} = \diag{
\begin{matrix}
	\Hn_{j,i}^{\left(1\right)}, \,
	\Hn_{j,i}^{\left(2\right)}, \,\Hn_{j,i}^{\left(3\right)}, \,
	\Hn_{j,i}^{\left(4\right)}
\end{matrix} } , \\[-0.5mm]
\Vn_{i} = \stack{
\begin{matrix}
	\Vn_{i}^{\left(1\right)}, \,
	\Vn_{i}^{\left(2\right)}, \,\Vn_{i}^{\left(3\right)}, \,
	\Vn_{i}^{\left(4\right)}
\end{matrix} },
\end{IEEEeqnarray*}
\\[-2mm]
where
$ {\Hn_{j,i}^{(p)} \Cmat{N\tau_p}{M\tau_p}}$, ${\Vn_{i}^{(p)} \Cmat{M\tau_p}{b}}$, ${\Hn_{j,i} \Cmat{N\tau}{M\tau}}\!\!$, and ${\Vn_i\Cmat{M\tau}{b}}$ are equivalent magnitudes for the global channel and precoding matrices, and matrix $\Gn_j^{(p)}$ is the signal space matrix for phase $p$ \cite{TAV_ppM1}.

We assume a delayed CSIT model, i.e. transmitters have only access to the previous phases CSI. Given the formulation, this means that the following channels
\begin{IEEEeqnarray*}{c}
 \{ \Hns{j,i}{\varrho}\}_{\varrho=1}^{p-1},\forall i,j,
\end{IEEEeqnarray*}
are available at the beginning of the phase $p$ at both BSs.

Finally, desired signals are separated from interference by applying a linear zero-forcing filter. Then, using standard derivations \cite{Cover} and assuming full-rank and generic channels, the achieved normalized DoF are given by:
\begin{IEEEeqnarray}{c}
\label{eq:achDoF}
\djin =  \frac1N\frac{b}{\tau}.
\end{IEEEeqnarray}

 \begin{figure}[]
\tikzstyle{every pin}=[font=\footnotesize,pin distance=0.6cm,inner sep=1pt]

\begin{tikzpicture}
\hspace{-2mm}
\begin{axis}[
  ymin=0.245,ymax=0.5,xmin=0.5,xmax=4.5,
  xmajorgrids,
   ymajorgrids,
   grid style={dashed, gray!30},
  xlabel style={at={(0.5,0.02)}},
  ylabel style={at={(0.07,0.43)},rotate=-90},
  width=1.05\linewidth, 
   xtick={0.25,0.5,1,1.333,2,3,2.6413,3.1557,3.5414,4}, %
   xticklabels={$\frac{1}{4}$,$\frac{1}{2}$,$1$,$\frac43$,$2$,$3$,$\roA$,$\roB$,$\roC$,$4$}, 
   ytick={0.125,0.25,0.333,0.4,0.4286,0.48,0.375},
   yticklabels={$\frac{1}{8}$,$\frac{1}{4}$, $\frac{1}{3}$,$\frac{2}{5}$,$\frac{3}{7}$,$\frac{12}{25}$,$\frac{3}{8}$},
   extra y ticks={},
   extra y tick labels={},
   x tick label style={xtick pos = left}, 
   y tick label style={ytick pos = left, yticklabel pos=left}, 
   extra y tick style={ytick pos = right, yticklabel pos=right},
  font=\scriptsize,
  ylabel=$\djin$,
  xlabel=${\rho = \nicefrac{M}{N}}$,
  legend style={at={(axis cs: 2.333,0.32)},anchor=north west,font=\scriptsize}
  ]
\newcommand{\K}{4}
  
  \poligon{(axis cs:2.6413,0.333) -- (axis cs:3,0.333) -- (axis cs:3,0.3803) -- (axis cs:2.6413,0.333)} 
  
 \poligon{(axis cs:3,0.375) -- (axis cs:4,0.375) -- (axis cs:4,0.4285) -- (axis cs:3.5414,0.4143) -- (axis cs:3.1557,0.4)
 -- (axis cs:3,0.3803) -- (axis cs:3,0.375)} 
  
 \poligon{(axis cs:4,0.4) -- (axis cs:4.495,0.4) -- (axis cs:4.495,0.4285) -- (axis cs:4,0.4285) -- (axis cs:4,0.4)}
  
  

  
  
  
  
  

        \pgfplotsset{
    plotoptsPrevIn/.style={color=blue, very thick, dotted},
    plotoptsPropIn/.style={color=orange, thick, dashed},
    plotoptsNoCSIT/.style={color=black, thick, dash pattern=on 3pt off 1pt on 1pt off 1.5pt},
    plotoptsPropUp2/.style={color=green, thick, dash pattern=on 3pt off 1pt on 1pt off 1.5pt},
}


  \addplot[red,  thick,domain=0:1/2]{x/2};
  \addlegendentry{Outer bound \cite{MAT}\cite{IBC_fullCSIT}};
   \addplot[forget plot, red,  thick,domain=1/2:3/4]{2*x/(2*x+3)};
   \addplot[forget plot, red,  ,  thick,domain=3/4:1]{1/3};
   \addplot[forget plot, red,  ,  thick,domain=1:4/3]{1/3*x};
 \addplot[forget plot, red,  thick,domain=4/3:3/2]{2*x/(3*x+2)};
 \addplot[forget plot, red,  thick,domain=3/2:2]{6*x/(11*x+3)};
  \addplot[forget plot, red,  ,  thick,domain=2:4.5]{12/25};
  
  \addplot[forget plot, red,  ,  thick,domain=1:6/5]{3*x};
  

  

  \addplot[plotoptsPropIn,domain=4:4.5]{3/7}; \label{plot:propin}  
       \addlegendentry{Proposed inner bound};   
       \addplot[forget plot,refstyle={plot:propin},domain=2.6413:3.1557)]{x^3/(x^3+3*x^2+3*x+8)};
       \addplot[forget plot,refstyle={plot:propin},domain=3.1557:1/2*(1+sqrt(37))]{3*x^2/(5*x^2+7*x+3)};
   \addplot[forget plot,refstyle={plot:propin},domain=1/2*(1+sqrt(37)):4]{9*x/(16*x+20)};


  \addplot[plotoptsPrevIn, domain=0:1/2]{x/2}; \label{plot:previn} 
      \addlegendentry{Previous inner bound \cite{MAT}\cite{IBC_dCSIT_ISIT14}};
    \addplot[forget plot,refstyle={plot:previn}, domain=0.5:1]{1/4}; 
    \addplot[forget plot,refstyle={plot:previn}, domain=1:2]{1/2*x/(x+1)}; 
    \addplot[forget plot,refstyle={plot:previn}, domain=2:3]{1/3}; 
    \addplot [forget plot,refstyle={plot:previn}, domain=3:4]{0.375}; 
    \addplot[forget plot,refstyle={plot:previn}, domain=4:4.5]{2/(\K+1)}; 
    
    
    \addplot[plotoptsNoCSIT,domain=1/2:4.5]{1/4};
  \addlegendentry{Without CSIT};

  
  

  

\end{axis}
\end{tikzpicture}

\vspace{-3mm} 
\caption{Normalized DoF per user inner and outer bounds for the 2-cell 2-user MIMO IBC with delayed CSIT for $\rho > \frac{1}{2}$. The optimal DoF for $\rho < \frac{1}{2}$ are attained without the need of CSIT, see \cite{IBC_fullCSIT}. The proposed scheme performance is depicted only for the region where previous inner bounds are outperformed, i.e. $\rho > \roA$.}
\vspace{-3mm}
\label{fig:mainResults} 
\end{figure}
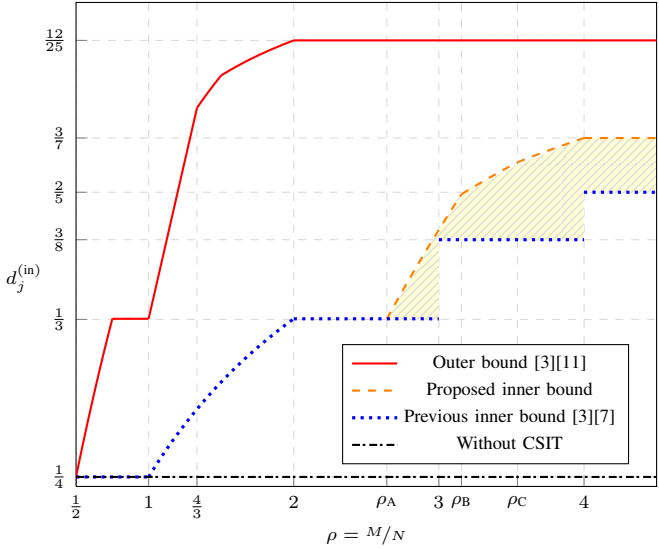   

\section{Main result}
\label{sec:MainResults}

 The next theorem summarizes the proposed DoF inner bound as a function of the antenna ratio $\rho = \frac{M}{N}$:

\vskip 2mm

%
%

\begin{theorem}
For the 2-cell 2-user MIMO IBC with delayed CSIT and antenna ratio $\rho$, \\[-4mm]
 \label{th:innerbound} 
\end{theorem}
%
\begin{IEEEeqnarray}{c}
 \djin = \left\{
 \begin{matrix}
  \frac{\rho^3}{\rho^3+3\rho^2+3\rho+8} & \roA \leq \rho < \roB \\ 
  \frac{3\rho^2}{5\rho^2+7\rho+3} & \roB \leq \rho < \roC \\ 
  \frac{9\rho}{16\rho+20} & \roC \leq \rho < 4 \\ 
  \frac{3}{7} & \rho \geq 4 
 \end{matrix}
 \right. \\[-3mm] \nonumber
 \label{eq:propDin}
\end{IEEEeqnarray}
normalized DoF per user can be achieved, where ${\roA \approx 2.6413}$, ${\roB \approx 3.2196}$, ${\roC \approx 3.5414}$. \\[-1mm]

\begin{IEEEproof}  
See Section \ref{sec:innerbound}. \\
\end{IEEEproof}   

Fig. \ref{fig:mainResults} summarizes current and previous DoF inner and outer bounds using delayed CSIT, as well as the optimal DoF when there is no CSIT for comparison purposes. Previous inner bounds are outperformed for $\rho > \roA$.

For $\rho<3$, previous inner bound curves correspond to the application of time division multiplexing to manage the inter-cell interference. This transforms the IBC into two independent 2-user BCs, where the MAT scheme \cite{MAT} for 2 users can be applied. On the other hand, \cite{IBC_dCSIT_ISIT14} showed that the DoF values $\frac{3}{8}$ and $\frac{2}{5}$ can be achieved for $\rho=3,4$\footnote{The DoF value for $\rho=4$ can also be achieved by applying the ideas reported in \cite{TAV} for the 4-user MISO IC.}, respectively. The DoF for the rest of settings are derived from the fact that increasing the number of transmit antennas cannot reduce the DoF.

We also depict the minimum between two trivial upper bounds as benchmark. First, assuming transmitter cooperation, we convert the IBC into a 4-user MIMO BC with $2M$ antennas at the transmitter, whose DoF were derived in \cite{MAT}. This bound is shown for $\rho > \frac43$. The second outer bound corresponds to assuming current CSIT \cite{IBC_fullCSIT}, and applies for $\rho<\frac43$.

\section{Proposed scheme}
\label{sec:innerbound}

Achievability of inner bounds in Theorem 1 is based on the transmission protocol described in this section. It is shown the precoding matrix design for the different phases, and that the required LCs of desired symbols are provided to each UE for correct decodability. Using this strategy forces some constraints over the different system parameters, deferred to \mbox{Appendix \ref{ap:OptProb}}, or the full paper \cite{TAV_IBC_fullPaper}. Therefore, the optimal parameters are obtained for each antenna setting by means of the following DoF maximization problem:
\begin{IEEEeqnarray}{c l}
 \underset{ \left\{b,S_1,S_2,S_3 \right\} \in \enters}{\text{maximize}} \quad  & \frac1N\frac{b}{4S_1+6S_2+4S_3+S_4} \label{eq:objfunc} \\[2mm]
s.t. & \Max{MS_1,4NS_1} \geq b \nonumber \label{eq:maxProblemaa} \\
     & MS_2 \geq NS_1 \nonumber \label{eq:maxProblemc} \\
     & MS_3 \geq 2NS_2 \nonumber \label{eq:maxProblemd} \\
     & N\pare{S_1+3S_2+6\Max{S_2,S_3}} \geq b \nonumber \label{eq:maxProbleme} 
\end{IEEEeqnarray}
where $S_4 = \Min{6S_2,\frac{b}{N} - S_1-3S_2} $. Due to space limitation, we address the specific case ${(M,N)=(4,1)}$, where $\frac{3}{7}$ DoF per user can be achieved by delivering $b=12$ symbols to each user along $\tau=28$ time slots, with $S_1=3$, $S_2=S_3=1$, and $S_4=6$. 


\subsection{Phase 1}
The first phase is divided in four rounds, each of them dedicated to transmit LCs of symbols of one user during 3 time slots. The precoding matrices used during this round are predefined before the communication. Consequently, after the first phase we have 3 LC of desired symbols, thus 9 additional LCs are required. Interestingly, those remaining LCs are distributed along the listening receivers, who observe 3 LCs of each non-desired set of symbols. These LCs will be the basis to align the interference during the next phases, while providing new LCs of desired symbols. They represent the overheard interference (OHI) for the first phase, denoted by  
\begin{IEEEeqnarray}{c}
 {\Tn{j,i}=\Hn_{j,c(i)}^{\left(1,i\right)}   \Vn_{i}^{\left(1,i\right)} \Cmat{3}{12}},
\end{IEEEeqnarray}
i.e. the signals intended to $\UE{i}$ and observed at $\UE{j}$, see Table \ref{tab:OHI} for the OHI collected at $\UE{1}$. Moreover, notice that all these $12$ distributed LCs can be assumed linearly independent with probability one due to the channel randomness. Finally, for better readibility we show the signal space matrix for $\UE{1}$:
\vspace{-4mm} \\
\begin{IEEEeqnarray}{c}
\def\arraystretch{1.4}
\Gn_1^{(1)} = \diag{ \Tn{1,1},\Tn{1,2},\Tn{1,3},\Tn{1,4}}
\label{eq:G1}
\end{IEEEeqnarray}



\subsection{Phase 2}
Users are served by pairs during this phase, with six \mbox{single-slot} rounds. The first phase OHI is exploited to deliver a LC of desired symbols aligned at the non-intended UE. Consider the round $r$ dedicated to $\UE{i}$ and $\UE{j}$, i.e. $\Ac^{(2,r)}\!\!=\!\{i,j\}$. Then, transmitted signals are designed such that
\begin{IEEEeqnarray*}{c}
        \RSpan{ 
        \Hns{j,c(i)}{2,r} \Vns{i}{2,r} } 
       \subseteq 
       \Tc_{j,i}, \label{eq:RIA_phase2b} \quad
  \RSpan{ 
        \Hns{i,c(j)}{2,r} \Vns{j}{2,r} } 
        \subseteq 
        \Tc_{i,j}, \label{eq:scC:RIA_phase2a}
\end{IEEEeqnarray*}
i.e. the signals obtained at each receiver are aligned with the previously received OHI. One simple solution using only delayed CSIT is given by
\begin{IEEEeqnarray}{c}
 \Vns{i}{2,r}  = \Sigmas{i}{2,r} \Tn{j,i} \label{eq:scC:desPhase2}, \quad
 \Vns{j}{2,r}  = \Sigmas{j}{2,r} \Tn{i,j},
 \label{eq:precPhase2}
\end{IEEEeqnarray}
where $\Sigmas{i}{2,r}, \Sigmas{j}{2,r}  \Cmat{4}{3} $ are some random full rank matrices ensuring the transmit power constraint. 

The OHI obtained during this phase is generally written as
\begin{IEEEeqnarray}{c}
 \tng{j,i}{k} = \hns{k,c(i)}{2,r} \Vns{i}{2,r} \Cmat{1}{12},
\end{IEEEeqnarray}
with $\RSpan{\tng{j,i}{k}}\subset \Tc_{j,i}$, and $r$ such that $\Ac^{(2,r)}=\{i,j\}$. In other words, $\tng{j,i}{k}$ is the LC of $\Tn{j,i}$ observed at $\UE{k}$. For instance, 
$ {\tng{4,3}{1} = \hns{1,2}{2,6} \Sigmas{3}{2,6} \Tn{4,3}}$ is the LC of $\Tn{4,3}$
observed at $\UE{1}$. It is worth pointing out that some of those terms are available coupled, as shown for $\UE{1}$ in Table \ref{tab:OHI}.

For ease of exposition, the signal space matrix for this phase at $\UE{1}$ is next shown:
\begin{IEEEeqnarray}{c}
\def\arraystretch{1.6}
\Gn_1^{(2)} = \left[ \,
\begin{matrix}
   \tng{2,1}{1} & \tng{1,2}{1} & \0 & \0 \\
   \tng{3,1}{1} & \0 & \tng{1,3}{1}  & \0 \\
   & \vdots & \vdots & \\
   \0 & \0 & \tng{4,3}{1}  & \tng{3,4}{1} 
   
\end{matrix} \, \right].
\label{eq:G2}
\end{IEEEeqnarray}
Finally, from the received signals represented by (\ref{eq:G1}) and (\ref{eq:G2}), each user is able to retrieve 3 new LCs of desired symbols by combining the current received signals and previous OHI.


\subsection{Phase 3}

Each of the four single-slot rounds of this phase is dedicated to a triplet of users. Precoding matrices are designed based on the OHI observed during the previous phase when the UEs act as listening users. In contrast, now the objective is twofold: \mbox{1) to} transmit LCs of desired signals, as in the previous phases, but also 2) to decouple the OHI from the second phase. On the one hand, transmitted signals are designed with the objective of being aligned with the previous received OHI, thus providing additional free of interference LC of desired signals. However, the generated interference will be aligned only at one of the receivers due to the effect of the two distributed transmitters, and the rest of receivers will obtain new LCs of interference, as explained next in this section. Then, at those receivers where not all the interference is aligned, the received signals will be useful to decouple their available OHI as shown for $\UE{1}$ in \mbox{Table \ref{tab:OHI}}.

\begin{table}[h]
\centering
\renewcommand{\arraystretch}{1.7}
\caption{OHI transmitted during each third phase round}
\begin{tabular}{C | C | C | C | C}
 r & \Ac^{(3,r)} & \Ac^{(3,r)}(1) & \Ac^{(3,r)}(2) & \Ac^{(3,r)}(3) \\
 \hline
 1 & \{1,2,3\} & \{ \tng{2,1}{3},\tng{3,1}{2} \}& \{ \tng{1,2}{3},\tng{3,2}{1} \} & \{ \tng{1,3}{2},\tng{2,3}{1} \} \\
 2 & \{1,2,4\} & \{ \tng{2,1}{4},\tng{4,1}{2} \}& \{ \tng{1,2}{4},\tng{4,2}{1} \} & \{ \tng{1,4}{2},\tng{2,4}{1} \} \\
 3 & \{1,3,4\} & \{ \tng{3,1}{4},\tng{4,1}{3} \}& \{ \tng{1,3}{4},\tng{4,3}{1} \} & \{ \tng{1,4}{3},\tng{3,4}{1} \} \\
 4 & \{2,3,4\} & \{ \tng{3,2}{4},\tng{4,2}{3} \}& \{ \tng{2,1}{3},\tng{3,1}{2} \} & \{ \tng{2,1}{3},\tng{3,1}{2} \} 
\end{tabular}
\label{tab:pieces3}
\end{table} 

 In order to illustrate these objectives, let us consider the precoding matrix design for the third round\footnote{Due to space limitation, we intentionally focus on this third round in order to make use of (\ref{eq:G2}). The design for the rest of rounds is easily derived according to Table \ref{tab:pieces3}.}, dedicated to users in $\Ac^{(3,3)}=\{1,3,4\}$:
\begin{IEEEeqnarray}{l}
\begin{matrix}
 \Vns{1}{3,3}  =  \sigmas{1}{3,3}\tng{3,1}{4} + \sigmas{2}{3,3}\tng{4,1}{3} , \\
 \Vns{3}{3,3}  =  \sigmas{1}{3,3}\tng{1,3}{4} + \sigmas{3}{3,3}\tng{4,3}{1} , \\
 \Vns{4}{3,3}  =  \sigmas{2}{3,3}\tng{1,4}{3} + \sigmas{3}{3,3}\tng{3,4}{1} ,
 \end{matrix}
 \label{eq:precPhase3}
\end{IEEEeqnarray}   
where $\sigmas{i}{3,r} \!\Cmat{4}{1}\!$ is some vector with random entries, thus the interference generated at $\UE{1}$ is:
\begin{IEEEeqnarray}{l l l}
 \inn_1^{\left(3,3\right)} & =\,\, & \hns{1,2}{3,3} \left( \Vns{3}{3,3} \mathbf{x}_3 + \Vns{4}{3,3} \mathbf{x}_4  \right) =  \label{eq:y33} \\ \nonumber
 & = \,\,& \hns{1,2}{3,3}  \Big( \sigmas{1}{3,3}\tng{1,3}{4} \xn_3
 +  \sigmas{3}{3,3}  \underline{\mathbf{a}} \Bigg[\begin{matrix} \xn_3 \\ \xn_4 \end{matrix}\Bigg] \!\!
 +  \sigmas{2}{3,3}\tng{1,4}{3} \mathbf{x}_4  \Big)
\end{IEEEeqnarray} \\[-3mm]
with $\underline{\mathbf{a}} = \Big[ \tng{4,3}{1}, \tng{3,4}{1} \Big]$. By exploiting the phase 2 OHI, see Table \ref{tab:OHI}, the terms proportional to $\mathbf{a}$ can be removed, while the rest of terms are aligned with the first phase OHI. Then, all the interference terms in (\ref{eq:y33}) are removed, and one additional LC of desired symbols is retrieved.
Unfortunately, the trick in (\ref{eq:y33}) can be done only if all the interference comes from the same transmitter, i.e. all the interference terms are observed through the same channel, as in a BC. To see this, consider the interference generated at this UE during the $(3,1)$th round:
\begin{IEEEeqnarray}{l l l}
 \inn_1^{\left(3,1\right)} & =\,  & \hns{1,1}{3,1} \left( 
 \sigmas{1}{3,1}\tng{1,2}{3} +  \sigmas{3}{3,1}\tng{3,2}{1} \right) \xn_2 \, + \label{eq:y31}  \\ \nonumber
 & +\, & \hns{1,2}{3,1} \left( 
 \sigmas{2}{3,1}\tng{1,3}{2} +  \sigmas{3}{3,1}\tng{2,3}{1} \right) \xn_3
\end{IEEEeqnarray} 

{\it Remark} 1: It is important to see that since $\tng{3,2}{1}$ and $\tng{2,3}{1}$ are known coupled (see Table \ref{tab:OHI})), they cannot be simultaneously canceled in (\ref{eq:y31}). However, this round allows $\UE{1}$ to decouple them, as expressed in $\mbox{Table \ref{tab:OHI}}$, to be exploited during the last phase.

Finally, the OHI for this phase is defined as follows:
\begin{IEEEeqnarray}{c}
 \Wn{j,i} = \hns{j,c(i)}{3,r} \Vns{i}{3,r} \Cmat{1}{12},
\end{IEEEeqnarray}
with $r$ defined such that $i \notin \Ac^{(3,r)}$. In other words, $\Wn{j,i}$ is the LC of symbols of $\UE{i}$ observed at $\UE{j}$ during the unique third round where $\UE{j}$ is a listening user. Note that ${\RSpan{\Wn{j,i}} \subset \mathsf{rspan}\big(\Stack\big(\tng{l,i}{k} ,\tng{k,i}{l}\big)\big)}$, where all indices take different values. For example, 
${\Wn{4,2} =  \hns{4,1}{3,1} \Vns{1}{3,1}  = \hns{4,1}{3,1} \left( 
 \sigmas{1}{3,1}\tng{1,2}{3} +  \sigmas{3}{3,1}\tng{3,2}{1} \right)}$,
 and its span lies on ${\mathsf{rspan}\Big(\Stack\big(\tng{1,2}{3} ,\tng{3,2}{1}\big)\Big)}$.

\subsection{Phase 4}
\label{sec:phase4}

The last phase consists of a single round of 6
 slots, during which all users are simultaneously served. In this case the precoding matrices transmit a rank-3 LC of symbols, thus there is some redundancy, that will be used not only to cancel the transmitted interference, but also to retrieve the desired LCs not properly decoded during the third phase. The precoding matrix design for this phase is generally written as:
 \\[-3mm]
\begin{IEEEeqnarray}{c}
  \mathbf{V}_{i}^{(4)}  =  \sum_{k=1}^{3} \sigmas{\Ic^i(k)}{4} \Wn{\Ic^i(k),i}, \label{eq:scC:precRIA}
\end{IEEEeqnarray}
\\[-2mm]
where $\sigmas{i}{4} \! \!\Cmat{4}{1}$, $\Ic^i = \{1,\dots,4\} \backslash \{i\} $, $\Ic^i(k)$ is the $k$th element of $\Ic^i$. Table \ref{tab:pieces4} shows the different LCs transmitted during this phase.
We next show how all the interference is removed at $\UE{1}$, detailed in rows 2, 3, 4. First, exploiting all the available OHI (\mbox{Table \ref{tab:OHI}}) most of the terms are removed. A concise description of this procedure may be found in Appendix \ref{ap:SSMoverview}. After this, $\UE{1}$ obtains
\begin{IEEEeqnarray}{c}
 {\Hnsg{1,1}{4}}{\bar \Vn_{1}^{(4)}} \xn_1 + 
 \underbrace{\Hns{1,1}{4} \sigmas{1}{4} \Wn{1,2}}_{\text{rank}\, 1 }\xn_2 + 
 \Hns{1,2}{4} \sigmas{1}{4} \big[ \Wn{1,3}, \Wn{1,4} \big] \Bigg[\begin{matrix} \,\xn_3\, \\ \xn_4 \end{matrix}\Bigg] \nonumber 
 \\ \label{eq:intermediateRxSignal}
\end{IEEEeqnarray}
\\[-5mm]
where $\Hnsg{1,1}{4}{\bar \Vn_{1}^{(4)}} \Cmat{6}{12}$ is a full rank matrix containing the desired signals received during phase 4, and rounds 1 and 2 of the third phase, i.e. the 5 LCs of desired signals not resolved up to this point due to interference. 
Now, using the OHI from the last third phase round $(\Big[ \Wn{1,2}, \Wn{1,3}, \Wn{1,4} \Big])$ the receiver is able to remove the remaining inter-cell interference terms. However, there remain some non-aligned intra-cell interference terms. Fortunately, 
$\Hns{1,1}{4} \sigmas{1}{4} \Wn{1,2} \Cmat{6}{12}$ has a \mbox{5-dimensional} orthogonal subspace where the signals can be projected. Then, all the interference is removed, and 5 LCs of desired signals can be retrieved from this phase.

\begin{table}[]
\centering
\renewcommand{\arraystretch}{1.7}
\caption{OHI transmitted during the last phase}
\begin{tabular}{C | C | C | C }
 i & \Wn{\Ic^i(3),i} & \Wn{\Ic^i(2),i} & \Wn{\Ic^i(1),i} \\
 \hline
 1 & \{ \tng{2,1}{3},\tng{3,1}{2} \} & \{\tng{2,1}{4},\tng{4,1}{2} \} & \{\tng{3,1}{4},\tng{4,1}{3} \} \\
 2 & \{ \tng{1,2}{3},\tng{3,2}{1}\} & \{\tng{1,2}{4},\tng{4,2}{1}\} & \{\tng{3,2}{4},\tng{4,2}{3} \} \\
 3 & \{ \tng{1,3}{2},\tng{2,3}{1}\} & \{\tng{1,3}{4},\tng{4,3}{1}\} & \{\tng{2,3}{4},\tng{4,3}{2} \} \\
 4 & \{ \tng{1,4}{2},\tng{2,4}{1}\} & \{\tng{1,4}{3},\tng{3,4}{1}\} & \{\tng{2,4}{3},\tng{3,4}{2} \}
\end{tabular}
\vspace{-2mm}
\label{tab:pieces4}
\end{table}

 Overall, the total number of free of interference LCs of desired symbols at the end of the communication is equal to
 $3 + 3 + 1 + 5 = 12 = b$,
thus $\djin = \frac{12}{28}=\frac{3}{7}$ DoF are achieved.

\section{Conclusion}
\label{sec:conclusion}

A four-phase protocol based on linear precoding and decoding is proposed for the 2-cell 2-user MIMO IBC with delayed CSIT. Assuming transmitters and receivers are equipped with $M$ and $N$ antennas, respectively, the proposed scheme improves best-known DoF inner bounds for the case ${\rho = \frac{M}{N}> 2.641}$. Further work is oriented to study other antenna settings, and scenarios with more cells or users per cell. {Moreover, since the optimal parameters for our scheme are derived by means of a maximization problem, some practical constraints may be introduced, e.g. maximum number of time slots, which is also an interesting line of future research.}  



\bibliographystyle{IEEEtran}
\bibliography{../../../papers/_referenciesMarc}


\appendices

\section{Derivation of the DoF maximization problem}
\label{ap:OptProb}


The proposed strategy forces some constraints over the different system parameters, as shown in the DoF maximization problem in (\ref{eq:objfunc}). Next, each of those constraints is derived for a general antenna setting. 
\vskip 1.5mm
1) {\it Transmit rank in phase 1}: During the first phase, $MS_1$ linear combinations of the $b$ symbols are transmitted using $M$ antennas, and during $S_1$ slots. Then, for linear decodability of the desired symbols, no more symbols than the number of transmit dimensions can be sent, thus we force $ M S_1 \geq b$.
\vskip 1.5mm
2) {\it Enough linear combinations in phase 1}: After the first phase, $NS_1$ linear combinations of each user's symbols are distributed along each UE. Note that no more fresh linear combinations of desired symbols will be introduced to the system, since the rest of phases consists on the retransmission of those linear combinations. Therefore, we force that $4NS_1 \geq b$, i.e. if each user have access to all $4NS_1$ linear combinations of its symbols, it should be able to linearly decode them.
\vskip 1.5mm
3) {\it Transmit rank in phase 2}: Since $NS_1$ linear combinations of overheard interference are retransmitted in the $S_2$ slots of each phase 2 round, we force $MS_2 \geq NS_1$, in order to ensure that the rank of transmitted overheard interference is preserved.
\vskip 1.5mm
4) {\it Enough linear combinations in phase 2}: After the second phase, $S_1+3S_2$ linear combinations of desired signals are obtained at each user, and $6S_2$ of desired symbols are distributed along the rest of receivers. Hence, we must have $N\pare{S_1+9S_2} \geq b$.
\vskip 1.5mm
5) {\it Transmit rank in phase 3}: Similarly to the first and third constraints, we force $MS_3 \geq 2NS_2$. 
\vskip 1.5mm
6) {\it Enough linear combinations for the whole communication}: At most, $N\pare{S_1+3 S_2+6 \Max{S_2,S_3}}$ linear combinations of desired symbols are delivered to each user. Then, we must have ${N\pare{S_1+3 S_2+\Max{S_2,S_3}}\geq b}$. This constraint generalizes the fourth constraint, which is then omitted.
\vskip 1.5mm

The problem in (\ref{eq:objfunc}) results from collecting all these constraints, and taking as objective function the number of symbols $b$ divided by the signal dimensions $N\tau$. 

Next, we show how the value of $S_4$ can be obtained in closed form. First, we have that $6S_2$ linear combinations of overheard interference are transmitted per UE during the last phase, thus we force:
\begin{IEEEeqnarray*}{c}
 {N\cdot 6S_2 } \leq N \cdot S_4.
\end{IEEEeqnarray*}
Moreover, after the third phase $N \pare{S_1+3S_2+S_3}$ linear combinations of desired symbols are provided to each UE, while the last phase delivers $N \pare{S_4-S_3}$ additional linear combinations to each of them. This last expression can be easily derived by taking into account the zero-forcing operation. Then, we must have
\begin{IEEEeqnarray*}{c}
 N \pare{S_1 + 3S_2 + S_3 + (S_4 - S_3) } \geq b \\
 N \pare{S_1 + 3S_2 + S_4  } \geq b
\end{IEEEeqnarray*}
Taking into account that no more than these two constraints involve $S_4$, we pick its minimum feasible, given by:
\begin{IEEEeqnarray*}{c}
 S_4 = \Min{6S_2,\frac{b}{N} - S_1-3S_2}  
\end{IEEEeqnarray*}
 
\section{$\UE{1}$ detailed perspective of \\ the precoding scheme}
\label{ap:SSMoverview}

For a better understanding of the precoding scheme, this appendix shows how the different signals are observed at $\UE{1}$, the different terms of overheard interference in \mbox{Table I} are obtained, and finally the interference is canceled. To this end, we write in the last page the signal space matrix for $\UE{1}$ during the first three phases, and the received signals during the last phase, together with \mbox{Table I} again, for the sake of reader's convenience. Dashed lines separate block rows of each phase.

From each UE perspective, the rounds of the first two phases can be classified in two classes: serving or listening round. During the serving rounds, extra linear combinations of desired signals are obtained, while the interference (if any) is aligned, thus can be removed. During the listening rounds, only interference is observed, to be used as overheard interference during the next phases. Note that after the second phase some of the overheard interference is {\it coupled}. For example, during the fourth round at $\UE{1}$, the overheard interference obtained is given by
\begin{IEEEeqnarray}{c}
 \yn_1^{(3,2)} = \inn_1^{(3,2)} = \tng{3,2}{1} \xn_2  + \tng{2,3}{1} \xn_3.
 \label{eq:OHI24}
\end{IEEEeqnarray}
Consequently, we denote the overheard interference known from this phase as $\Big[\tng{3,2}{1},\tng{2,3}{1} \Big]$. The rest of overheard interference terms acquired at $\UE{1}$ is specified in \mbox{Table I}.

In contrast to the first two phases, three different types of rounds may be defined during the third phase. In case of $\UE{1}$, the third and fourth rounds are serving and listening rounds, respectively, following the definition above. However, the first and second rounds correspond to an intermediate situation, denoted as mixed rounds. In those cases, desired signals are received but combined with interference terms that cannot be removed. This is because they are transmitted from different BSs, i.e. observed through different channels, and because the overheard interference is known coupled. For example, consider the first round, where the received signal contains the term
\begin{IEEEeqnarray*}{c}
 \hns{1,1}{3,1} \sigmas{3}{3,1}\tng{3,2}{1} \xn_2  + \hns{1,2}{3,1} \sigmas{3}{3,1}\tng{2,3}{1} \xn_3.
\end{IEEEeqnarray*} 
$\UE{1}$ only knows (\ref{eq:OHI24}), i.e. it does not know each of the terms individually, thus it will not be able to remove them together. Then, what the receiver can do is to at least remove one of them in order to decouple these two overheard interference terms. For this reason, \mbox{Table I} shows that after this round the available overheard interference is $\{\tng{3,2}{1},\tng{2,3}{1}\}$ instead of $\Big[\tng{3,2}{1},\tng{2,3}{1} \Big]$, i.e. the overheard interference is {\it decoupled}.

And finally, the last phase consists of a single round. The received signals at $\UE{1}$ during its 6 slots are explicitly written in (\ref{eq:y14apendix}) in the last page. For some matrix $\mathbf{A}$ (different for each case), the subspaces occupied by interference  during this phase can be classified in three classes:
\begin{itemize}
 \item $\An \tng{1,i}{k}, \forall k\neq 1,i\neq 1$: All those terms can be removed using the first phase overheard interference, since by definition $\RSpan{\tng{j,i}{k}}\subset \Tc_{j,i}$.
 \item $\An \tng{j,i}{1}, \forall j\neq 1,i$: All those terms can be removed using the second phase overheard interference, which are now decoupled thanks to the mixed rounds of the third phase.
 \item Other terms that cannot be individually removed.
\end{itemize}

After suppressing the two first classes of interference terms, the remaining interference may be written as
\begin{IEEEeqnarray*}{c}
\Hns{1,1}{4} \Sigmas{4}{4} \Wn{1,2} \xn_2 + 
        \Hns{1,2}{4}   \Sigmas{4}{4}
        \left[ \Wn{1,3} ,  \Wn{1,4} \right] 
        \Bigg[\begin{matrix} \, \xn_3 \, \\ \xn_4 \end{matrix}\Bigg]
 \end{IEEEeqnarray*}
where the last term can be removed using the overheard interference obtained from the last round of the third phase. However, there remains some intra-cell interference, as shown in (\ref{eq:intermediateRxSignal}). Those remaining terms, in this case containing symbols intended to $\UE{2}$ are removed through zero-forcing concepts, as explained in Section \ref{sec:phase4}, after equation (\ref{eq:intermediateRxSignal}).

\begin{figure*}  
\centerline{\includegraphics[width=0.92\linewidth]{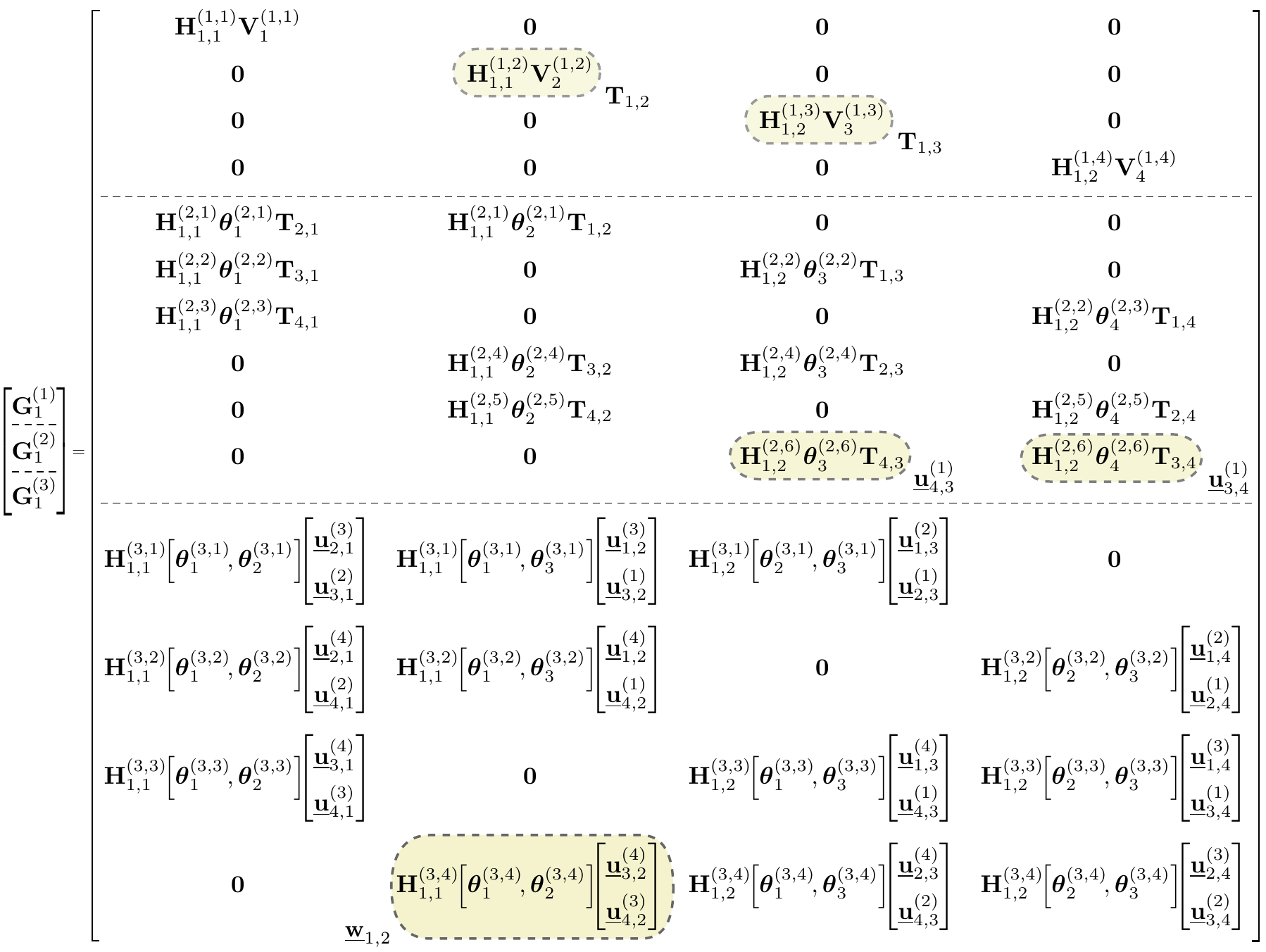}}   
\end{figure*}

\begin{table*}
\hspace{-2mm}
\scalebox{0.88}{
\def\arraystretch{2}
\begin{tabular}{|C | C | C | C | C | C | C | C | C | C | C | C | C | C | C |}
\hline
\multicolumn{4}{|c|}{Phase 1} & \multicolumn{6}{c|}{Phase 2} &
\multicolumn{4}{c|}{Phase 3} & \multicolumn{1}{c|}{Phase 4} \\
\hline 
 \rotatebox{90}{ \hspace{2.5mm}\rotatebox{-90}{  \{1\} }} & 
 \rotatebox{90}{ \hspace{2.5mm}\rotatebox{-90}{  \{2\} }} & 
 \rotatebox{90}{ \hspace{2.5mm}\rotatebox{-90}{  \{3\} }} & 
 \rotatebox{90}{ \hspace{2.5mm}\rotatebox{-90}{  \{4\} }} & 
\rotatebox{90}{ \,\{1,2\} }  & 
\rotatebox{90}{ \,\{1,3\} } &
\rotatebox{90}{\,\,\,\{1,4\} } &
\rotatebox{90}{\,\,\,\{2,3\}} &
\rotatebox{90}{\,\,\,\{2,4\}} &
\rotatebox{90}{\,\,\,\{3,4\}} &
\rotatebox{90}{\,\{1,2,3\}} &
\rotatebox{90}{\,\{1,2,4\}} &
\rotatebox{90}{\,\{1,3,4\}} &
\rotatebox{90}{\,\{2,3,4\}} &
\rotatebox{90}{\!\{1,2,3,4\} } \\
\hline
     & \Tn{1,2} & \Tn{1,3} & \Tn{1,4} &
 & & & \Big[ \tng{3,2}{1}, \tng{2,3}{1} \Big] & 
 \Big[ \tng{4,2}{1}, \tng{2,4}{1} \Big] &
 \Big[ \tng{4,3}{1}, \tng{3,4}{1} \Big] & 
 \tng{3,2}{1}, \tng{2,3}{1} &
 \tng{4,2}{1}, \tng{2,4}{1}  & &
 \Big[ \Wn{1,2},\Wn{1,3},\Wn{1,4} \Big] & \\[0.8mm]
\hline
\end{tabular} 
}
\end{table*}

\begin{figure*} 
 \scalebox{0.95}{
 \begin{minipage}{\linewidth}
    \begin{IEEEeqnarray}{r c l}
  \yn_1^{(4)}  = \,\, & \Hns{1,1}{4}  & \label{eq:y14apendix}
        \left( \big(\Sigmas{1}{4} \Wn{4,1}   +	\Sigmas{2}{4} \Wn{3,1}+ \Sigmas{3}{4} \Wn{2,1}\big)\xn_1 + \big( \Sigmas{1}{4} \Wn{4,2}  + \Sigmas{2}{4} \Wn{3,2}+ \Sigmas{4}{4} \Wn{1,2}\big) \xn_2 \right) + \\ \nonumber
        & \Hns{1,2}{4} &
        \left( \big(\Sigmas{1}{4} \Wn{4,3}   +	\Sigmas{2}{4} \Wn{3,3}+ \Sigmas{4}{4} \Wn{1,3}\big)\xn_3 + \big( \Sigmas{2}{4} \Wn{3,4}  + \Sigmas{3}{4} \Wn{2,4}+ \Sigmas{4}{4} \Wn{1,4}\big) \xn_4 \right) 
        \\[2mm] \nonumber
   = \,\, &\Hns{1,1}{4}  &
        \left( \Big( \Sigmas{1}{4} \Hns{4,1}{3,1} \sigmaU{1}{3,1}{2}{3,1}{2,1}{3}{3,1}{2}   +
	\Sigmas{2}{4} \Hns{3,1}{3,2} \sigmaU{1}{3,2}{2}{3,2}{2,1}{4}{4,1}{2}	+
	\Sigmas{3}{4} \Hns{2,1}{3,3} \sigmaU{1}{3,3}{2}{3,3}{3,1}{4}{4,1}{3}\Big) \xn_1 \right. + \\  \nonumber
	&& \left. \quad \!\! \Big( \Sigmas{1}{4} \Hns{4,1}{3,1} \sigmaU{1}{3,1}{3}{3,1}{1,2}{3}{3,2}{1}   +
	\Sigmas{2}{4} \Hns{3,1}{3,2} \sigmaU{1}{3,2}{3}{3,2}{1,2}{4}{4,2}{1}	+
	\Sigmas{4}{4} \Hns{1,1}{3,4} \sigmaU{1}{3,4}{2}{3,4}{3,2}{4}{4,2}{3} \Big) \xn_2 \right) + \\ \nonumber
	& \Hns{1,2}{4}  &
        \left( \Big( \Sigmas{1}{4} \Hns{4,2}{3,1} \sigmaU{2}{3,1}{3}{3,1}{1,3}{2}{2,3}{1}  +
	\Sigmas{3}{4} \Hns{2,2}{3,3} \sigmaU{1}{3,3}{3}{3,3}{1,3}{4}{4,3}{1}	+
	\Sigmas{4}{4} \Hns{1,1}{3,4} \sigmaU{1}{3,4}{3}{3,4}{2,3}{4}{4,3}{2} \Big) \xn_3 \right. + \\  \nonumber
	&& \left. \quad \!\! \Big( \Sigmas{2}{4} \Hns{3,2}{3,2} \sigmaU{2}{3,2}{3}{3,2}{1,4}{2}{2,4}{1}   +
	\Sigmas{3}{4}  \Hns{2,2}{3,3} \sigmaU{2}{3,3}{3}{3,3}{1,4}{3}{3,4}{1} 	+
	\Sigmas{4}{4} \Hns{1,1}{3,4} \sigmaU{2}{3,4}{3}{3,4}{2,4}{3}{3,4}{2} \Big) \xn_4 \right)
  \end{IEEEeqnarray}
 \end{minipage}
  }
 \end{figure*}

\end{document}